\documentclass[aps,prb,epsfig,twocolumn,floats,showpacs,amsmath,amssymb]{revtex4}
\usepackage{graphicx}
\usepackage{hyperref}
\begin{document}

\title{Magnetic field induced two-channel Kondo effect in multiple quantum dots}

\author{Konstantin Kikoin}
\affiliation{Physics Department, Ben-Gurion University,
Beer-Sheva, 84105, Israel}
\author{Yuval Oreg}
\affiliation{Department of Condensed Matter Physics, Weizmann
Institute of Science, Rehovot, 76100, Israel}
\date{\today}

\begin{abstract}
We study the possibility to observe the two channel Kondo physics in
multiple quantum dot heterostructures in the presence of magnetic
field. We show that a fine tuning of the coupling parameters of the
system and an external magnetic field may stabilize the two channel
Kondo critical point. We make predictions for behavior of the
scaling of the differential conductance  in the vicinity of the
quantum critical point, as a function of magnetic field, temperature
and source-drain potential.
\end{abstract}

\maketitle
\newpage
\section{Introduction}

Simple models for non-Fermi liquids, such as the multi-channel
Kondo models, have attracted recently much theoretical interest.
The predicted thermodynamic properties as well as transport
properties are markedly different from the properties of a Fermi
liquid. For example the conductance is predicted to approach its
asymptotic zero temperature value as the square root of the
temperature.

In this work we study a system built of a small quantum dot with
even number of electrons coupled to a large dot and free leads. We
explore the possibility to observe two channel Kondo physics in it
by a fine tuning of the magnetic field and the couplings between
the small quantum dot, the leads and the large dot.

In the first experimental realization of the {\emph {single channel
Kondo}} effect in planar quantum dot \cite{Goldhaber98} transport
was measured through a small dot containing odd number of electrons,
which may be described by a local spin impurity model. At low
temperature the spin is screened by the conduction electrons (of the
single channel), a Kondo resonance is formed and the conductance
through the dot approaches (in the symmetric dot case) to the
universal value $G_0=2 e^2/(2 \pi \hbar)$ with a Fermi liquid $T^2$
law. Soon after the experimental realization of the single channel
Kondo effect it was understood theoretically that in the presence of
magnetic field \cite{Pustilnik00} a \emph {single channel
\underline{even} Kondo} resonance may be formed in a dot with an
even number of electrons. For example, when there are two electrons
in a two-level dot the relevant states are a singlet state $\left|
00 \right\rangle$ with total spin $S=0$ and energy $E_S$; and a
triplet of states, $\left|1 S_z\right\rangle$ with $S=1$ and
$S_z=-1,0,1$ and energies $E_{T,S_z} = E_T+E_z S_z$, ($E_z = g \mu_B
h$ is the Zeeman energy). If $E_S < E_T$ than the application of
magnetic field reduces the energy of the triplet state with $S_z=1$
until at $h_c \equiv (E_T-E_S)/g \mu_B$ the states $\left|1
S_z=1\right\rangle$, $\left| 00 \right\rangle$ are degenerate. The
doublet of the degenerate state may be considered as isospin with
which the conduction electrons form the Kondo state. Following these
theoretical predictions the even Kondo effect was observed
experimentally in vertical dots \cite{Sasaki00} and
nanotubes\cite{Nygard00}.

It is important to appreciate that although the screening of the
spin in both the single-channel Kondo effect and the
single-channel-even Kondo effect is a non trivial many body
phenomenon, eventually at low temperature,  the spin is fully
screened and both effects exhibit conventional Fermi liquid
behavior. In contrast, when the dot has a doubly degenerate
ground-state and there are two independent channels that equally
screen the spin, an over-screening occurs and a non trivial many
body two channel Kondo effect is formed. This many body state has
both thermodynamic and transport properties that do not follow the
predictions of the Fermi liquid theory.

The main difficulty in realizing a physical system that materializes
the two channel Kondo model is in creating two separate channels
that equally screen the spin. In conventional setups an electron
from one channel that hops on the dot may hop to the other channel
and causes mixing between the channels. This mixing lead eventually
to two``eigen channels" with one channel coupled stronger than the
other one. The channel with the stronger coupling fully screen the
spin and the other channel is decoupled, and we thus have again the
single channel Kondo case. It was suggested \cite{Oreg03}  to
overcome this mixing problem by using a large quantum dot as an
additional channel.  Then, the free leads form one channel and the
large dot forms the second channel (cf. Fig~\ref{fg:model}). The
channels do not mix as transfer of electrons between them will
charge the large dot. This suggestion for the realization of the
\emph {two channel Kondo effect} was realized
experimentally~\cite{Potok06}.

In this work we combine the two channel Kondo effect with the even
Kondo effect. We will show that it is possible to tune the
magnetic field, the coupling strengthes of the small dot to the
large dot and to the free leads in such a way that the two channel
fixed point is stabilized.

 The  paper is arranged as follows: in section \ref{se:model} we
introduce a model that describes a small dot with even number of
electrons connected to a large dot and free reservoirs. In
subsections \ref{subse:Symmetry} and \ref{subse:Projection} we
introduce a few simplifying assumptions concerning the symmetry of
the dot -- lead couplings, and   define the appropriate vector
operators $\bf P$ which describe the effective spin built from the
singlet and one of the triplet states in the presence of
accidental degeneracy induced by a magnetic field. Then in
subsection \ref{subse:SW} we integrate out the high energy fields
using a generalized Schrieffer -- Wolff transformation until we
obtain the effective low temperature Hamiltonian
[Eq.~(\ref{eq:hanizo})], which in general case demonstrates both
channel and exchange anisotropy. As a special case we consider a
situation when the small dot is split to two small dots.  In
section \ref{se:RG} we analyze  the flow of the parameters using a
Renormalization Group approach and discuss the conditions to
observe the two channel Kondo physics. We find that in spite of
the complicated form of Hamiltonian (\ref{eq:hanizo}), the fixed
point for two channel Kondo effect may be achieved by fine tuning
the setup parameters. In section \ref{se:Experimental} we discuss
possible experimental realizations and observations of the
predictions of the model. Some details of the calculations are
relegated to appendices that appear after the summary.

\section{Model}
\label{se:model}

 Generalizing the idea formulated in Ref. \onlinecite{Oreg03}, we
study a system consisting of a small dot $d$ with two  discrete
energy levels labeled by index $m=1,2$ coupled with left $L$ and
right $R$ leads and a large dot $C$, by means of tunneling
amplitudes $t_{mL}, t_{mR}, t_{mC}$, respectively. (We have assumed
here that the coupling constants do not depend on the state indices
of the leads and the large dot.) The key feature of the model is
that the dot $C$ is sufficiently large so that the single particle
mean level spacing $\delta$ in this dot is small enough to make its
spectrum quasi continuous, $\delta \ll T_K$, whereas the Coulomb
blockade energy $E_C$ is large enough to fix the number of electrons
${\cal N}_C$, i.e. to prevent occupation of this dot by extra
electron injected from the leads or from the small dot $d$.

\begin{figure}[h]
\includegraphics[width=8cm]{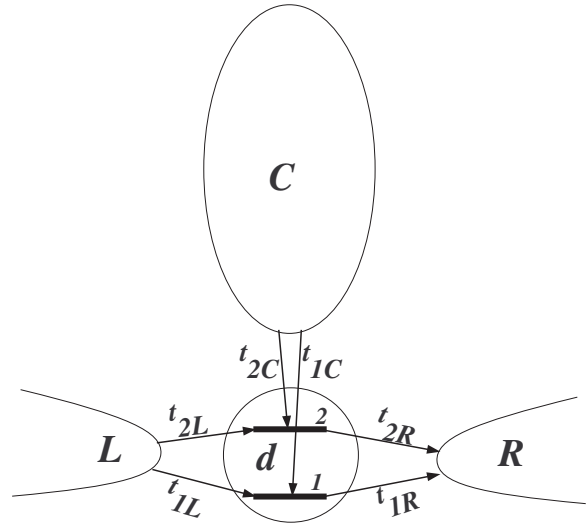}
\caption{A small dot $d$ with even number of electrons coupled to a
large dot $C$ and two leads $L$ and $R$. There are two levels in dot
$d$. The charging energy $U$, in the small dot $d$ is larger than
the charging energy $E_C$ in the large dot $C$.}\label{fg:model}
\end{figure}

We model the system (cf. Fig.~\ref{fg:model}) by the Hamiltonian
${\cal H}={\cal H}_d+{\cal H}_C+{\cal H}_{\rm leads}+{\cal H}_{\rm
tun}$, with
\begin{eqnarray}
&& H_d=\sum_\Lambda E_\Lambda |\Lambda\rangle\langle\Lambda|\equiv
\sum_\Lambda E_\Lambda X^{\Lambda \Lambda} \nonumber \\
&&  H_C=\sum_{k \sigma} \epsilon_{Ck \sigma} {\hat n}_{C k \sigma}+
E_C{\cal N}_C^2\;\;\; ({\cal N}_C=\sum_{k \sigma} {\hat n}_{C k
\sigma})
\nonumber\\
&& H_{\rm lead}=\sum_{B=L,R}\sum_{k\sigma}\epsilon_{k}
a^\dagger_{B k \sigma}a^{}_{B k \sigma} \label{eq:Hamiltonian}\\
&& H_{\rm tun}=\sum_{k\sigma}\sum_{m=1,2}\sum_{A=B,C}\left(
t_{mA}d_{m\sigma}^\dagger a^{}_{A k \sigma} + h.c.\right).
\nonumber
\end{eqnarray}

Here $a^\dagger_{B k \sigma}, a^{\phantom{\dagger}}_{B k \sigma}$
are the creation and annihilation operators in the free left and
right leads $B = L, R$, $a^\dagger_{C k \sigma},
a^{\phantom{\dagger}}_{C k \sigma}$ are the creation and
annihilation operators of the large dot $C$, ${\hat n}_{Ck\sigma}
= a^\dagger_{Ck}a^{\phantom{ \dagger}}_{Ck\sigma}$, and the set of
states $\left\{|\Lambda\rangle\right\}$ denote the exact many body
eigenstates of the isolated small dot $d$. In our notation the
eigen energies include not only the strong Coulomb blockade but
also the Zeeman term, which do not influence the electrons in the
free lead and the electrons in the large dot as by assumption they
have a constant density of states. The operators $X^{\Lambda
\Lambda } \equiv
\left|\Lambda\right\rangle\left\langle\Lambda\right|$ are the
Hubbard operators\cite{Hubbard65} (see Appendix
\ref{App:Spinalgebra}).

\subsection{Symmetry properties of the tunneling matrix elements
$t_{mi}$} \label{subse:Symmetry}

When there is only one level in the dot, it is possible to perform a
unitary transformation of the electron creation and annihilation
operators in the left and right leads so that only one linear
combination is coupled to the small dot, while the other one is
decoupled~\cite{Glazman88}. In the generic case it is not possible
to perform such rotation for the multilevel dot.\cite{Silva02}
However, if the tunneling matrix elements do not depend on $k$ and
\begin{equation}
\label{eq:t-symmetry}
 \frac{t_{L1}}{t_{R1}}=\frac{t_{L2}}{t_{R2}} = \alpha \equiv e^{i
 \varphi} \tan \theta
\end{equation}
we can define:
\begin{eqnarray}
\label{eq:psi-phi}
a_{\Psi k \sigma} = e^{i \varphi} \left[\phantom{-} \cos\theta a_{Rk\sigma} + \sin\theta a_{Lk\sigma}\right] \nonumber \\
a_{\Phi k \sigma} = e^{i \varphi} \left[         -  \sin\theta a_{Rk\sigma} + \cos\theta a_{Lk\sigma}\right]
\end{eqnarray}
which diagonalize the left-right sector of the Hamiltonian. To
obtain $H_{\rm lead}$ in terms of $a_{\Psi k \sigma}$ and $a_{\Phi
k \sigma}$ we simply have to substitute in
Eq.~(\ref{eq:Hamiltonian}) $B=\Phi,\Psi$, but the tunneling
Hamiltonian is now:
\begin{equation}
H_{\rm tun}=\sum_{k\sigma}\sum_{m=1,2}\sum_{A=\Psi,C}\left(
t_{mA}d_{m\sigma}^\dagger a_{A k \sigma} + h.c.\right).
\label{htus}
\end{equation}
with
$$
t_{1\Psi}= t_{1R}/\cos^2\theta,\;\;t_{2\Psi}= t_{2R}/\cos^2\theta.
$$

The $\Phi$-channel is decoupled from the dot. Through out the
paper we will assume that the symmetry
relation~(\ref{eq:t-symmetry}) holds.

\subsection{Projection to $\left| S \right\rangle \;\; \left| T1 \right \rangle$ subspace}

\label{subse:Projection}

The interesting physics will occur in a certain occupation of the
small dot $d$. To understand what is the relevant subspace, we
first assume that the strong Coulomb energy $U$ fixes the number
of electrons in the small dot to be even (say ${\cal N}_d=2$).
When the single particle level spacing in the small $d$-dot,
denoted by $\delta$, obey the inequality $t_{mi}\ll \delta \ll U$,
we may further assume that only two single particle levels
$\epsilon_{m},\; m=1,2$ in ${\cal H}_d$ are involved in the
tunneling.  So one can model the small dot $d$ by a two-level dot
with two electrons.

 We denote the single particle levels of the dot by  $\epsilon_{1}$
 and $\epsilon_2$ and assume that
 $\epsilon_2>\epsilon_1$.
 There are six possible states in this system, but we will neglect
 for
 simplicity the possibility of double occupation of level 2 and
 the singlet state of electrons occupying different levels.
 We thus remain with a subset of four lowest states:
 a spin singlet state, $|S\rangle$, with energy $E_S$ when two electrons occupy level 1; and a
 spin triplet, $|T\rangle$, with energy $E_T$, when one electron occupy level 1
 and the other level 2.

In terms of the  creation operators with
$d_{1\sigma}^\dagger,~d_{2\sigma}^\dagger$ of electrons with spin
$\sigma$ at levels 1 and 2 respectively, the singlet state is given
by
\begin{equation}\label{sing}
|S\rangle =d^\dagger_{1\uparrow}d^\dagger_{1\downarrow} \left|{\rm
vac} \right\rangle;
\end{equation}
and the three triplet components by
\begin{eqnarray}
\left|T_1        \right\rangle &=&d^\dagger_{1\uparrow}  d^\dagger_{2\uparrow}
\left|{\rm vac} \right\rangle, \nonumber\\
\left|T_0        \right\rangle &=&\frac{1}{\sqrt2}
\left(d^\dagger_{1\uparrow}d^\dag_{2\downarrow}+d^\dag_{1\downarrow}d^\dag_{2\uparrow}\right)
|{\rm vac}\rangle,\nonumber\\
\left|T_{\bar 1} \right\rangle
&=&d^\dagger_{1\downarrow}d^\dagger_{2\downarrow} \left|{\rm vac}
\right\rangle.
\label{trip}
\end{eqnarray}

The two-electron levels $E_\Lambda$ are the eigenvalues of
Hamiltonian $H_d$ truncated in the way described above:
\begin{eqnarray}
E_S&=&2\epsilon_{1}+U \label{level}\\
E_T&=&\epsilon_{1}+\epsilon_{2}+U-J. \nonumber
\end{eqnarray}
We further assume that the exchange splitting $J$ obeys the
equation $\epsilon_2-\epsilon_1  > J > 0$. In this case the
singlet has a lower energy in the absence of magnetic field.

Next, in accordance with the standard approach
\cite{Oreg03,Pustilnik04}, we neglect the finite level spacing
$\delta_{C}$ in the large dot $C$, but suppose that the Coulomb
blockade $E_C$ fixes the number of electrons, ${\cal N}_C$, in the
large dot. The interval $\delta_{C}$ denotes the lower limit of the
energy of electron-hole excitation in dot $C$.

\begin{figure}[h]
\includegraphics[width=6cm]{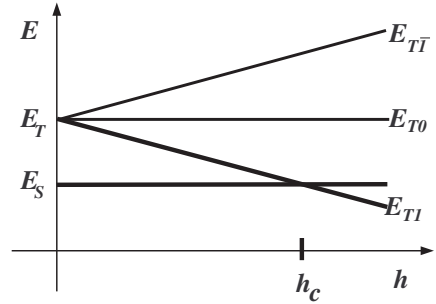}
\caption{Accidental degeneracy of a spin multiplet induced by Zeeman
splitting.}\label{fg:AD}
\end{figure}

The Kondo tunneling regime is ineffective in our system unless
$\delta = \epsilon_2 -\epsilon_1 <  J$. This is a special situation,
which may be achieved in some cases by tuning the gate voltage or
playing with diamagnetic shift.\cite{Kogan03,Zumbuhl04} We study
here a general case of singlet ground state and appeal to the
mechanism of accidental degeneracy induced by an in-plane external
magnetic field $h$, which occurs when the Zeeman effect compensates
the singlet--triplet splitting.
\cite{Pustilnik00,Nygard00,Giulanoshort0001} This mechanism is
illustrated in Fig. \ref{fg:AD}. In our model the degeneracy between
the states $E_{T1}$ and $E_S$ occurs when the condition
\begin{equation}
    \delta-J=E_Z
\end{equation}
is satisfied. Here $E_Z=g\mu_B h$ is the Zeeman splitting energy.

To find out the possibility of a Kondo effect induced by an external
magnetic field, one should derive the effective spin Hamiltonian
containing relevant variables. Various procedures for the
description of this phenomenon were discussed in
Refs.~\onlinecite{Pustilnik00,Giulanoshort0001}. We will use the
formalism of dynamical symmetry group, offered and elaborated in
Refs.~\onlinecite{KikoinShort0104}. According to this approach
briefly described in Appendix~\ref{App:Spinalgebra}, the spin
degrees of freedom in case of accidental degeneracy of the states
$\left|S\right\rangle$ and $\left|T1\right\rangle$ are described by
the vector ${\bf P}$ with components
\begin{equation}
\begin{array}{rcccc}
P^+ &=&\left| T1 \right\rangle \left\langle S  \right| &\equiv& f^\dagger_\uparrow f_\downarrow, \nonumber \\
P^- &=&\left| S  \right\rangle \left\langle T1 \right| &\equiv& f^\dagger_\downarrow f_\uparrow, \nonumber \\
P^z &=& \frac{1}{2}\left(\left|T1\right\rangle \left\langle T1
\right| - \left|S \right\rangle \left\langle S \right|\right)
&\equiv& \frac{1}{2}\left(f^\dagger_\uparrow f_\uparrow -
f^\dagger_\downarrow f_\downarrow \right) .\label{pse}
\end{array}
\end{equation}

The operator ${\bf P}$ obeys conventional commutation relations of
momentum operator. It can be treated as an effective spin operator
describing transversal and longitudinal spin excitations in a
quantum dot occupied by two electrons under condition of accidental
degeneracy illustrated by Fig.~\ref{fg:AD}. We therefore may
introduce an effective spin operator whose creation and annihilation
operators with components $\sigma=\uparrow, \downarrow$ are denoted,
respectively, by $f^\dagger_\sigma, f_\sigma$.

\subsection{Integration of high energy states and the Schriefer-Wolff transformation}
\label{subse:SW}

To obtain the effective low energy Hamiltonian one has to integrate
in the renormalization group sense the states with high energy.
There are few energy scales here: the charging energy of the small
dot, $U$, the charging energy of the large dot $E_C$ and the precise
position of the many body levels in the dot. We will assume that
this integration have renormalized the original parameters of
Hamiltonian~(\ref{eq:Hamiltonian}) and will concentrate on a
situation when only the state $\left| 00 \right\rangle$  and $\left|
11 \right\rangle$ and single-electron states of the dot are relevant
for virtual hopping between  $d$ and $L$, $R$ and $C$.
 We will also assume that
the scale of the energy considered is smaller than $E_C$ so that
transition between $L$ and $R$ to $C$ (via $d$) are excluded.

 After concentrating on the two states of the dot we are in a
 position to perform the Schrieffer-Wolff \cite{Schriefer66}
 transformation. The Schrieffer-Wolff transformation leads to
 terms of the form:
 \begin{equation}
 \label{eq:SW}
 \sum_{\sigma,\sigma'=\uparrow,\downarrow \; A, A' = \Psi, C;} J^{s s'}_{A A'}
     c^\dagger_{A \sigma} c^{\phantom{\dagger}}_{A' \sigma'}
     f^\dagger_{\sigma'}
 f^{}_{\sigma}.
 \end{equation}
In the general case $J^{\sigma \sigma'}_{A A'}$ is a tensor of
rank $6$. However, due to the symmetry
assumption~(\ref{eq:t-symmetry}) it reduces to a rank 4 tensor.
For example the term $J^{\uparrow \uparrow}_{CC}$ describes a
virtual transition from the triplet state to the large dot $C$ and
back. This transition occurs when an electron hops from the large
dot $C$ on the small dot $d$ and back, or when one of the
electrons that build the triplet state of dot $d$ hops on dot $C$
and back. According to our previous assumptions the former process
is negligible and there are two contributions to the latter
process. In the first an electron from level 1 hop on dot $C$ and
back and in the second electron from level 2 hop on dot $C$ and
back. Similar processes should be taken into account
for~$J^{\sigma,\sigma'}_{\Psi\Psi}$.

We have therefore:

\begin{equation}
\label{eq:JCC} J^{\uparrow \uparrow}_{AA} =
\sum_{m=1,2}\frac{\left| t_{mA}\right|^2}{\epsilon_{m}+U-J-E_Z/2}.
\end{equation}
Here the exchange integral is estimated for the intermediate state
with the electron escaped from the dot $d$ to the Fermi level
which is supposed to be the same for the leads and the dot $C$,
$\epsilon_{F\uparrow}=\epsilon_{F\downarrow}=0$. The SW
Hamiltonian is derived under the condition:

\begin{equation}\label{compa}
\epsilon_{2}-J-E_Z=\epsilon_{1}.
\end{equation}
Using this degeneracy condition, we derive
\begin{eqnarray}
\label{eq:JCC1}
 J^{\downarrow \downarrow}_{AA} &=&
\frac{\left| t_{1A}\right|^2}{\epsilon_{1A}+U+E_Z/2} \\
 J^{\uparrow \downarrow}_{AA} &=& \left(J^{\downarrow \uparrow}_{AA}\right)^*=
\frac{ t_{1A} t_{2A}^*}{\epsilon_{1}+U+ E_Z/2}.\nonumber
\end{eqnarray}
Notice that due to the presence of magnetic field the $SU(2)$
symmetry is broken and $J^{\sigma \sigma'}_{AA}$ depend on
$\sigma$ and~$\sigma'$. In the spirit of usual SW approximation,
one may neglect the small differences in the denominators of above
exchange integrals and estimate them as
\begin{eqnarray}
J^{\uparrow \uparrow}_{AA}
&=&\frac{\left|t_{1A}\right|^2+\left|t_{2A}\right|^2}{\epsilon_{1}+U}\label{eq:JCC1}\\
J^{\downarrow \downarrow}_{AA} &=& \frac{\left|
t_{1A}\right|^2}{\epsilon_{1}+U},~~~ J^{\uparrow \downarrow}_{AA} =
\frac{ t_{1A} t_{2A}^*}{\epsilon_{1}+U}.\nonumber
\end{eqnarray}

Due to the charging energy $E_C$, transitions between the $\Psi$,
and $C$ are excluded so that $J_{\Psi C}^{\sigma \sigma'}=0$, in
addition $J_{\Phi A}=J_{A \Phi}=0,\; A=\Phi,\Psi,C$ since it was
decoupled from the dot due to the symmetry
assumption~(\ref{eq:t-symmetry}).

Rearranging the terms in (\ref{eq:SW}), we may write the Hamiltonain
in terms of the components of the vector $\bf P$ and the conduction
electron spin $\mathbf{\sigma(x=0)}$ (In the general case $H_{SW}$
has a more complicated form)
\begin{eqnarray}
&& H_{\rm SW}=\sum_{A=\Psi,C} \frac{J_{\perp A}}{2}\left(P^+
\sigma^-_A +P^-
\sigma^+_A\right)\label{eq:Hsw}\\
&& + \sum_A J_{zA}P^z \sigma_A^z
      + \sum_A {J_{0A}}P^z \sigma_A^0 +H_{\rm potential} \nonumber
\end{eqnarray}
where $H_{\rm potential}$, describes potential scattering of
conduction electrons without spin flips, ${\sigma_A^0}_{\alpha
\beta}\equiv n_A/2=\sum_{kk^\prime} c^\dagger_{A k \alpha} c_{A k
\beta} \delta_{\alpha \beta}/2$ and
\begin{eqnarray}
J_{\perp A}&=& 2 J_{AA}^{\uparrow \downarrow} \nonumber \\
J_{z A}    &=& J_{AA}^{\uparrow \uparrow}+ J_{AA}^{\downarrow \downarrow}\nonumber \\
J_{0 A}    &=& J_{AA}^{\uparrow \uparrow} - J_{AA}^{\downarrow
\downarrow}
\end{eqnarray}

The total effective Hamiltonian $H_{\rm eff}$  contains an
additional term $\propto P_z n_A$, which does not exist in the
standard two channel Kondo Hamiltonian. \cite{Nozieres80} In the
presence of magnetic field we may write it as:
\begin{eqnarray}\label{eq:hanizo}
H_{\rm eff}=\sum_A  \frac{J_{\perp A}}{2}( P^+ \sigma^-_A+  P^+
\sigma^-_A) +
J_{zA}P^z \sigma^z_A\nonumber\\
+ \sum_A J_{0A}P^z (n_A- \bar n_A) + B_{Z}P^z +H_{\rm lead}.
\end{eqnarray}
Here the first two terms describe an effective {\it multichannel}
Kondo Hamiltonian, which is anisotropic both in spin and channel
variables and includes effective magnetic field $B_{Z}$. This
``field" is small in comparison with the Zeeman field $E_Z$ (see
Fig. \ref{fg:AD}),
\begin{equation}\label{eq:be}
B_Z=\frac{E_{T1}-E_S}{2} +\sum_A \frac{J_{0A}}{2} \bar n_A.
\end{equation}
Here $\bar n_A$ is the average electron density in channel $A$. It
is seen from (\ref{eq:JCC1}) that an exchange anisotropy arises from
the fact that there are two tunnel matrix elements $t_{1A}$,
$t_{2A}$ in $H_{\rm tun}$ [see Eq.~(\ref{htus})], which enter the
longitudinal and transversal exchange constants in different
combinations.

\begin{figure}[h]
\includegraphics[width=8cm]{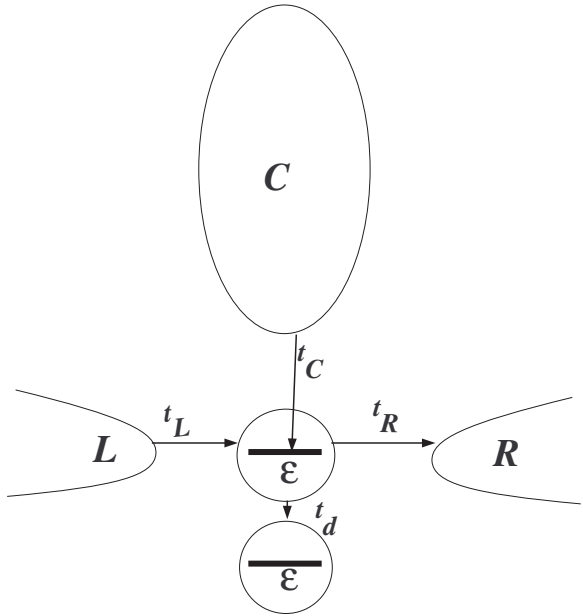}
\caption{Experimental setup with a small dot split into two
equivalent wells.}\label{fg:Expsetup}
\end{figure}

The exchange anisotropy may be eliminated from $H_{SW}$ if the
experimental setup is modified in accordance with Fig.
\ref{fg:Expsetup}. In this modification the small dot is split into
two valleys coupled by tunneling element $t_d$. Each of these
valleys is assumed to be occupied by odd number of electrons. If the
highest unoccupied levels are equivalent, then the levels
$\epsilon_{1,2}$, which enter the Hamiltonian $H_{d}$ are
\begin{equation}\label{ddot}
\epsilon_{1,2} = \epsilon \mp t_d;
\end{equation}
the singlet-triplet splitting in Eqs.~(\ref{level}) is determined by
the indirect exchange $J'=2t_d^2/U$ instead of direct exchange $J$;
and the SW procedure described above gives an effective Hamiltonian
in which only one tunnel matrix element, $t_C$, enters all
parameters $J^{\sigma\sigma'}_{CC}$ (\ref{eq:JCC1}). In contrust to
the model of Fig. \ref{fg:model}, with the split dot we obtain the
standard two-channel Hamiltonian with isotropic exchange coupling
(and no $J_{0A}$) in an effective magnetic field,
\begin{equation}\label{be2}
H_{\rm eff}= \sum_A J'_A{\bf P} \cdot {\mbox{\boldmath
$\sigma$}}_A+B'_{Z}P^z
\end{equation}
where  $B'_{Z}$ is the effective Zeeman field given by the first
term in Eq.~(\ref{eq:be}).

\section{Renormalization group equations for effective
Hamiltonian} \label{se:RG}

To derive the flow equations for exchange vertices in a
two-channel Hamiltonian, the conventional scaling procedure is
used \cite{Anderson70,Fowler71} and 3-rd order diagrams  that
contain loops are included.\cite{Nozieres80}
 We start with the setup of Fig.
\ref{fg:Expsetup} which is described by the  Hamiltonian (\ref{be2})
with isotropic exchange. In this case the system of flow equations
has the form
\begin{equation}\label{scal3}
\frac{dj_a}{d \eta}= -j_a^2 +j_a(j_a^2+j_b^2)
\end{equation}
where $b=\Psi$ if $a=C$ and vice versa, $j_a=\rho_{a}J_{a}$,
$\rho_a$ is the corresponding density of states on the Fermi level
and $\eta= \ln D$ is the current scaling variable with initial
value $D=D_0$.
 In accordance with the Nozieres-Blandin scenario,
the flow diagram $(j_\Psi, j_C)$ for this system has a weak
coupling fixed point, which is unstable against channel
anisotropy.

The system (\ref{scal3}) is written for the case of $B_{Z}=0$. It is
known that the external magnetic field is a relevant parameter for
the conventional two-channel Kondo effect.\cite{Affleck92} In our
case the two-channel regime arises only at certain magnetic field
$h=h_c$ (Fig. \ref{fg:AD}) and, similar to the effect of an external
magnetic field in conventional two channel Kondo system, the
deviation $h-h_c$ from this critical value is detrimental for the
two-channel physics.

The system of flow equations for the anisotropic Hamiltonian
(\ref{eq:hanizo}) with dimensionless coupling constants
$j_a^\iota$ is more complicated. In this case the third-order
corrections in the r.h.s. of scaling equations contains the sums
$j_a^\iota\sum_b\sum_{\kappa\neq \iota}(j_b^\kappa)^2$, where
$a,b$ are the channel indices and the indices $\iota,\kappa$ mark
the Cartesian components $x,y,z$ in spin space. Besides,
additional scaling equations should be written for the integrals
$j_{0a}$.

It is easily seen that the vertices $j_a^0=\rho_a J_{a0}$ remain
unrenormalized in 2-nd order, and the third-order corrections enter
with plus sign in the r.h.s. of corresponding scaling equations.
Thus, the parameters $j_a^0$ scale into zero. As to the anisotropic
third order terms in scaling equations for $j_a^i$, it is known from
predictions of conformal field theory \cite{Affleck93}, that the
exchange anisotropy is irrelevant for two-channel regime. So we
conclude that the flow diagram for the two-channel Hamiltonian
(\ref{eq:hanizo}) with anisotropic exchange coupling retains the
finite isotropic fixed point and only the channel anisotropy is
relevant.

\section{experimental realization and consequences}
\label{se:Experimental}

\subsection{Setup}

The even charge Kondo effect can be realized experimentally in a
setup similar to the one that was suggested to the odd two-channel
Kondo effect \cite{Oreg03}. In the ordinary  setup (that was
realized recently experimentally \cite{Potok06}) two
noninteracting leads (L and R) and a large dot C are attached to a
single-level small dot d. In the setup considered here we need
more then one level on the small dot and require also to fulfil
certain relations between the coupling
constants~(\ref{eq:t-symmetry}). In addition, even when these
symmetries are fulfilled, we still need to tune the magnetic field
to be at a degeneracy point between the singlet and the triplet,
and fine tune the dot-channel coupling so that $J_\psi=J_c$. In
practice this would be rather complected as many parameters should
be tuned.

In a setup with two-valley double dot shown in
Fig.~\ref{fg:Expsetup} only one tunnel constant enters all
exchange integrals. In this case the two-level dot $d$ is formed
by two single particle dots. Two dot levels $\epsilon_{1,2}$ are
given by Eq. (\ref{ddot}) and the single particle states of the
small dot $d$ are the symmetric and antisymmetric wave functions
of single dots. As a result the coupling constants to the two
levels are equal and the ratio (\ref{eq:t-symmetry}) is fulfilled
authomatically. This condition facilitates the fine tuning
procedure.

\subsection{Expected results}

In order to identify the even channel Kondo effect one has first
to identify the Kondo case when the finite reservoir $C$ is open.
This can be found by tuning the magnetic field, and the levels of
the small dot so that $B_Z =0$. Next when dot $C$ is formed and
tunneling of electrons between $L$ and $R$ to $C$ is forbidden,
one has to tune the coupling constants so that $J_\Psi=J_C$.

We will assume in the following the the coupling to lead $L$ is much
bigger than the coupling to lead $R$. Then as in the regular two
channel Kondo effect \cite{Oreg03,Potok06} we expect that when
$J_\psi=J_C$ differential conductance between the left and right
leads $g(V,T)$ (with $V$ the voltage and $T$ the temperature) will
follow the 2CK scaling law \cite{Pustilnik04,Affleck93}:
\begin{equation}
\label{eq:dofcon1} \frac{2}{g_0} \frac{g(0,T)-g(V,T)}{\sqrt {\pi T
/T_{K2}}} = Y\left(\frac{\left|e V\right|}{\pi T}
 \right)
\end{equation}
with $g_0 = e^2 /h$ and the scaling function $Y(x)$ is given in
Appendix \ref{app:Y}.

While $J_\psi \ne J_C$, the scaling law is:
\begin{equation}
\label{eq:ScalingCurves}
 \frac{1}{g_0} \frac{g(0,T)-g(V,T)}{\left(\pi T /T_{\Delta}\right)^2} = \text{sign}\left(\Delta\right)
\frac{3}{2}\left(\frac{e V}{\pi T}\right)^2
\end{equation}
with $\Delta = J_C-J_\Psi \ll J_\Psi, J_C$ and $T_\Delta =
\Delta^2/J_\Psi^4 T_K$.

\subsubsection{Effect of Magnetic Field}
Unlike the single channel case,  the magnetic field is a relevant
parameter in the two channel case. We expect that the effective
magnetic field in the even two channel Kondo case will operate in a
way very similar to the operation of the ordinary magnetic field in
the two channel Kondo case \cite{Pustilnik04}. Notice though that
the effective magnetic field $B_Z$ [see Eq. (\ref{eq:be})] is
controlled by the real external magnetic field, by the position of
the two levels and by the coupling constants.

Namely for $\sqrt{T_K T } \ll B_Z \ll B_\Delta =
\left(\Delta/J_\Psi^2 \right) T_K$ we have, for the linear
conductance $g(V=0,T,B)$ the relation:
\begin{equation}
\label{eq:Bz} \frac{1}{2 g_0} \frac{g(0,T,0)-g(0,T,B_Z)}{\left(
B_Z/B_\Delta\right)^2} = -\text{sign}\left(\Delta\right).
\end{equation}
For $B_\Delta\ll B_Z \ll T_K$ the conductance is given by the
Bethe-ansatz solution~\cite{Pustilnik04,Andrei95}

\begin{equation}
\label{eq:Bz1} \frac{g(0,T,B_Z)}{2 g_0}-1/2 = a \
\text{sign}\left(\Delta\right) \frac{B_\Delta}{B_Z} - b
\frac{B_Z}{T_K} \log \frac{T_K}{B}.
\end{equation}
with $a$ and $b$ of ${\cal O}(1)$.

\section{Summary}
We studied the possibility to observe the two channel Kondo effect
in multiple dots hetero-structures with even occupation in the
presence of magnetic field. Like in case of single channel Kondo
tunneling \cite{Pustilnik00}, we have found that magnetic field,
which is as a rule detrimental for odd occupation Kondo regime, may
induce the two-channel Kondo effect provided the Zeeman splitting
compensates the exchange gap between singlet ground state and the
state with spin projection oriented parallel to magnetic field. The
effective spin of the small quantum dot in this case is 1/2 despite
of the even occupation, so that the two channels provided by the
large dot and the leads are sufficient to overscreen the dot spin.
The exchange anisotropy which arises in presence of magnetic field
disappears because of additional degeneracy in case when the small
dot is split into two identical wells (Fig. \ref{fg:Expsetup}). The
two channel Kondo quantum critical physics may be stabilized by fine
tuning of the parameters (i.e. coupling constants and external
magnetic field).

\section{Acknowledgement}
We acknowledge useful discussion with Eran Sela, Ileana Rau and
David Goldhaber-Gordon. The research was supported by DIP, ISF and
BSF foundations.

\appendix
\section{Spin algebra for doubly occupied quantum dot}
\label{App:Spinalgebra}
 In case of strong Coulomb blockade, the
low-energy part of the spectrum of complex quantum dots with even
occupation consists of a system of singlet/triplet pairs (in the
absence of the orbital symmetry, which may impose the Hund scheme
of level occupation). These levels may be tuned by external
magnetic field and gate voltages, and the accidental degeneracy
arises as a result of level crossing. Under these conditions the
notion of dynamical symmetry is extremely
useful.\cite{KikoinShort0104} This symmetry characterizes not only
the ground state of a system, but also the transitions between the
eigenstates of quantum dot within a given energy interval (in our
case this interval is determined by the Kondo temperature $T_K$).
The simplest object, which possesses a dynamical symmetry is the
S/T multiplet.

It is convenient to describe the eigenstates of a given Hamiltonian
and the intra-multiplet transitions by the Hubbard operators
$X^{\Lambda_1\Lambda_2}=|\Lambda_1\rangle\langle \Lambda_2|$. These
operators obey  the obvious multiplication rule
$X^{\Lambda_1\Lambda_2}X^{\Lambda_3\Lambda_4}=X^{\Lambda_1\Lambda_4}\delta_{\Lambda_2\Lambda_3}$,
from which the commutation relations may be easily
derived.\cite{Hubbard65}  The dynamical symmetry of S/T multiplet is
characterized by two vectors ${\bf S}$ and ${\bf R}$ defined as
\begin{eqnarray}
S^+ & = & \sqrt{2}\left(X^{u0}+X^{0d}\right),~ S_z =
X^{uu}-X^{dd}.
 \label{m.1} \\
R^+ & = & \sqrt{2}\left(X^{uS}-X^{Sd}\right), ~R_z =
-\left(X^{0S}+X^{S0}\right). \nonumber
\end{eqnarray}
These two vectors obeying $o_4$ spin algebra generate $SO(4)$
symmetry group of spin rotator.\cite{KikoinShort0104} The
dynamical symmetry of spin rotator is realized in Kondo tunneling
near the point of accidental S/T degeneracy.

As is known, \cite{Pustilnik00,KikoinShort0104} the Zeeman
splitting $E_Z=g\mu_B h$ reduces $SO(4)$ symmetry of CQD, when the
external magnetic field induces the resonance $E_Z\approx
\Delta_{ex}.$ In this case one deals with a system of two quasi
degenerate levels $E_S, E_{Tu}$, whereas two other components of
triplet are quenched at low energy $\epsilon\ll E_Z$. The
dynamical symmetry of this two-level system is described by a set
of operators $\{X^{SS},X^{uu},X^{Su},X^{uS}\}$, which form a
closed $SU(2)$ algebra.  To show this, one may introduce a pair of
vectors ${\bf P}$, ${\bf Q}$  instead of (\ref{m.1}). The
components of these vectors are defined as
\begin{eqnarray}\label{m.2}
P^+ &=&X^{uS},~~~P^z=\frac{1}{2}(X^{uu}-X^{SS}),\nonumber \\
Q^+ &=&X^{0d},~~~Q^z=\frac{1}{2}(X^{00}-X^{dd}).
\end{eqnarray}
These two vectors realize  the mathematical possibility of
factorization of the spin rotator symmetry group into a direct
product, $SO(4)=SU(2)\times SU(2)$, so that each vector transforms
as an effective spin 1/2. Only first of these vectors is relevant
in the resonance condition  $|E_Z - \Delta_{ex}|\sim T_K$, where
$T_K$ is the Kondo temperature, which characterizes cotunneling
involving effective "spin" ${\bf P}$ with usual commutation
relations $[P^i,P^j]= i\varepsilon_{ijk}P^k$, where $i,j,k$ are
cartesian components of this vector.

\section{The scaling function $Y(x)$}
\label{app:Y} The function $Y(x)$ is given by:
\begin{eqnarray}
\label{eq:I2}
 Y(x) &=&-1 -\int_0^1 du \frac{3}{\pi}\frac{1}{\sqrt{u(1-u)^3}}
      \nonumber \\
&\times & \left[ \frac{|\log u|\sqrt{u}}{1-u} F(u)\cos(x \log
u)-1\right],
\end{eqnarray}
$$
F(u)=\frac{1 -\sqrt{u}}{2}E\left(\frac{-4\,{\sqrt{u}}}
      {{\left( {\sqrt{u}-1} \right) }^2}\right) \label{eq:I1F}
$$
with $E(x)$ the complete elliptic function.

It has the asymptotic form:

\begin{equation}
\label{eq:Ylimits}
Y(x) \approxeq \left\{ \begin{array}{cc}
c \ x^2 &  \text{ for } x \ll 1   \\
 \frac{3} {\sqrt{\pi}}  \sqrt{x} -1 & \text{ for } x \gg 1   \\
\end{array} \right.
\end{equation}
with $c=0.748336$.


\begin{thebibliography}{21}

\bibitem{Goldhaber98} D. Goldhaber-Gordon, H. Shtrikman, D.
Mahalu, D. Abusch-Magder, U. Meirav, and M.A. Kastner, Nature,
\textbf{391}, 156 (1998).

\bibitem{Pustilnik00} M. Pustilnik,  Y. Avishai, and K. Kikoin,
 Phys. Rev. Lett. \textbf{64}, 1756{2000}.

\bibitem{Sasaki00}
S. Sasaki, S.D. Franceschi,  J.M. Elzerman,  W.G. van~der Wiel, M.
Eto,  S. Tarucha, and  L.P. Kouwenhoven,  Nature \textbf{405}, 764
{2000}).

\bibitem{Nygard00}J. Nygard,  D.H. Cobden,  and P.E.  Lindelof, Nature \textbf{408}, 342 {2000}.

\bibitem{Oreg03} Y. Oreg and D. Goldhaber-Gordon, Phys. Rev. Lett. \textbf{90},  136602 {2003}.

\bibitem{Potok06} R.M. Potok,  I.G. Rau,  H. Shtrikman, Y. Oreg , and D. Goldhaber-Gordon
   {2006} {submitted for publication}.

\bibitem{Hubbard65} J. Hubbard, Proc. Roy. Soc. A \textbf{285}, 542  {1965}.

\bibitem{Glazman88}
L.I. Glazman and M.E. Raikh, Sov.Phys. Lett. \textbf{47}, 452
{1988}, [Pis'ma Zh. Eksp. Teor. Fiz. {\bf 47}, 378 (1988)].

\bibitem{Silva02} A. Silva, Y. Oreg, and Y. Gefen, Phys. Rev. B \textbf{66}, 195316 {2002}.

\bibitem{Pustilnik04} M. Pustilnik, L. Borda, L.I. Glazman, and J. von Delft, Phys. Rev. B \textbf{69},
   115316 {2004}.

\bibitem{Kogan03} A. Kogan, G. Granger, M.A. Kastner, D. Goldhaber-Gordon, and H. Shtrikman,
   Phys. Rev. B \textbf{67}, 113309 {2003}.

\bibitem{Zumbuhl04} D.M. Zumb\"uhl, C. Marcus, M.P. Hanson, and A.C. Gossard, Phys. Rev. Lett.
  \textbf{93}, 256801 (2004).

\bibitem{Giulanoshort0001} D. Giuliano and A. Tagliacozzo, Phys. Rev. Lett. {\bf84}, 4677
  (2000); Pustilnik and L. Glazman, Phys. Rev. Lett. {\bf 85}, 2993 (2000);
  {\it ibid.}, Phys. Rev. B {\bf64}, 045328 (2001).

\bibitem{KikoinShort0104}
K. Kikoin and Y. Avishai, Phys. Rev. Lett. {\bf86}, 2090 (2001);
  {\it ibid.}, Phys. Rev. B {\bf65}, 115329 (2002), T. Kuzmenko, K. Kikoin and
  Y. Avishai, Phys. Rev. B {\bf 69} 195109 (2004).

\bibitem{Schriefer66} J.R. Schrieffer and P.A. Wolff, Phys. Rev. \textbf{149},
  491 {1966}.

\bibitem{Nozieres80} P. Nozieres and A. Blandin, J. Phys. (Paris) \textbf{41},
  193 {1980}.

\bibitem{Anderson70} P.W. Anderson, J. Phys. C.: Solid State \textbf{3},
  2436 {1970}.

\bibitem{Fowler71} M.Fowler and A. Zawadowski, Solid St. Commun. \textbf{9},
  471 {1971}.

\bibitem{Affleck92} I. Affleck, A.W.W. Ludwig, H.B. Pang and D.L. Cox,
  Phys. Rev. B \textbf{45}, 7918 {1992}.

\bibitem{Affleck93} I. Affleck and A.W.W. Ludwig, Phys. Rev. B \textbf{48},
  7297 {1993}.

\bibitem{Andrei95} N. Andrei and A. Jerez, Phys. Rev. Lett. \textbf{74},
  4507 {1995}.
\end{thebibliography}
\end{document}